# Flow-induced Density Fluctuation assisted Nucleation in Polyethylene

Xiaoliang Tang,[1] Junsheng Yang,[1,2] Fucheng Tian,[1] Tingyu Xu,[1] Chun Xie[1] and Liangbin Li[1][1]

[1]National Synchrotron Radiation Lab and CAS Key Laboratory of Soft Matter Chemistry, Anhui Provincial Engineering Laboratory of Advanced Functional Polymer Film, University of Science and Technology of China, Hefei, 230026, China
[2]Computational Physics Key Laboratory of Sichuan Province, Yibin University, Yibin, 644000, China

**Abstract**

The nucleation process of polyethylene under quiescent and shear flow conditions are comparatively studied with all-atom molecular dynamical simulations. At both conditions, nucleation are demonstrated to be two-step processes, which, however, proceed via different intermediate orders. Quiescent nucleation is assisted by local structure order coupling conformational and local rotational symmetric orderings, while flow-induced nucleation is promoted by density fluctuation, which is a coupling effect of conformational and orientation orderings. Flow drives the transformation from flexible chains to rigid conformational ordered segments and circumvents the entropic penalty, which is the most peculiar and rate-limited step in polymer crystallization. Current work suggests that flow accelerates nucleation in orders of magnitude is not simply due to flow-induced entropic reduction of melt as early models proposed, which is mainly attributed to the different kinetic pathway via conformational/orientational ordering – density fluctuation – nucleation.

* Correspondence author: lbli@ustc.edu.cn

# INTRODUCTION

Flow induced crystallization (FIC) is a non-equilibrium phase transition relevant to annual processing of millions metric tons of semi-crystalline polymeric materials globally, which has been attracted great attention for decades [1-3]. Imposing flow can accelerate nucleation rate in orders of magnitude and change the morphology from isotropic spherulitic to oriented shish-kebab structures [4-11], which not only raises processing efficiency but also enhances mechanical, thermal, optical and other properties of final products. To account flow-induced nucleation (FIN), the most widely recognized entropy reduction model (ERM) states that flow orientates or stretches polymer chains and consequently lowers nucleation barrier of crystal [12]. Through incorporating entropic reduction $\Delta S_f$ due to chain orientation and stretch, the nucleation barrier under flow is expressed as $\Delta G_f^* = \Delta G_q^* + T\Delta S_f$, where $\Delta G_q^*$ is the nucleation barrier at quiescent from classic nucleation theory (CNT) [13]. To account the new structure and morphology of nuclei, recently a modified ERM is proposed, which considers flow-induced free energy changes of both the initial melt and the final crystal nuclei. All these models for FIN are essentially based on CNT and assume that the nucleation kinetic pathway is the same at flow and at quiescent conditions. CNT states that the transition from liquid to crystal is a one-step process, which enjoys a great success at qualitative level but is unfortunately hard to predict nucleation rate quantitatively. As ERM confines itself to the one-step framework of CNT, naturally one would not expect that FIN can be quantitatively interpreted by ERM.

The one-step framework of CNT is challenged by two-step nucleation models in recent decades [14-20], in which either density fluctuation [21-23] or bond-orientational order fluctuation [24,25] is proposed to assist crystal nucleation. Those two-step models emphasize the existence and the importance of intermediate states (or precursor) during nucleation. With molecular dynamic simulation, recently we show that crystal nucleation of polyethylene (PE) is a two-step process assisted by a local structure order (LSO, denoted as $O_{CB}$) at quiescent condition [26], where $O_{CB}$ is an order parameter coupling conformational order and rotational symmetry order of neighboring atoms but without the requirement of density or orientational orders. Comparing to spherical atoms and small molecules, how flexible chain transform into rigid conformational ordered segment (COS) is the most peculiar rate-limited step in polymer crystallization, which can be overcome by $O_{CB}$ with the cooperative effect of COS with rotational symmetry at quiescent condition. As flow can induce conformational order like *gauche-trans* or coil-helix transitions [27,28] and align chain segments in parallel, different structural intermediates may emerge, resulting in different kinetic pathways of nucleation as comparing to that at quiescent condition. Indeed, the emergence of non-crystalline shish with density contrast to matrix melt is well documented in FIC experiments[29-31], while at quiescent density fluctuation prior to nucleation is still a controversial issue lacking of solid evidence, suggesting that nucleation under flow and quiescent conditions may follow different kinetic pathways.

In this work, with all-atom molecule dynamic simulation we comparatively study the nucleation processes of PE under quiescent and shear conditions (see APPENDIX). With the order parameter $O_{CB}$, the same as our former work[25], we identify a local ordered structure with symmetry similar to hexagonal of PE lattice (denoted as $H$-$O_{CB}$) and orthorhombic crystal (denoted as $O$-$O_{CB}$), while density is expressed by Voronoi volume. By analysis the simulation results, we observe that nucleation of PE crystal at quiescent and at flow all follow two-step processes but with different intermediate orders. Quiescent nucleation is assisted by $H$-$O_{CB}$ fluctuation, while FIN undergoes different kinetic pathway mediated via density fluctuation. This suggests that flow enhanced nucleation rate in orders of magnitude may be mainly due to the new nucleation pathway via density fluctuation rather than the entropic reduction of melt stated in early models.

## RESULTS

**Quiescent condition.** The system goes through a 20 ns NPT ensemble process at 390 K and the evolutions of $H$-$O_{CB}$ and $O$-$O_{CB}$ structures are shown in the Fig. 1(a), which are calculated using the $O_{CB}$ parameter defined in our former work[26] (also see APPENDIX). Even though the force-field is different in this work, similar phenomenon is observed that the clusters with $O_{CB}$ value matching hexagonal symmetry ($H$-$O_{CB}$) form stochastically in the early stage, while it takes an incubation time of about 7 ns for orthorhombic nuclei ($O$-$O_{CB}$) to emerge. Fig. 1(b) plots the evolutions of the average Voronoi volumes (high value corresponds to low density) of melt and nucleation atoms, respectively. Note nucleation atoms are first labeled in $O$-$O_{CB}$ clusters at 20 ns and then calculate their Voronoi volume during simulation from 0 to 20 ns, during which these atoms can be in either melt or $H$-$O_{CB}$ state before the formation of $O$-$O_{CB}$ structure. An obvious increase of density is accompanied with the formation of orthorhombic nuclei ($O\text{-}O_{CB}$), while the average density of $H\text{-}O_{CB}$ clusters remains the same as the matrix melt, indicating that local structure order (LSO) of $H\text{-}O_{CB}$ does not couple with density fluctuation. Slices of the simulation system with thickness of 50 nm at the early (before 7 ns) and the final states are shown in Figs. 1 (c) and (d) (only $O_{CB}$ structures are presented), respectively. Carbon atoms colored in red and yellow correspond to $O\text{-}O_{CB}$ crystals while the blue and green atoms are for $H\text{-}O_{CB}$ clusters. The $H\text{-}O_{CB}$ clusters are dynamic in nature and grow in size with time. After the incubation time, the $O\text{-}O_{CB}$ nuclei emerge inside of the $H\text{-}O_{CB}$ domains. Figs. 1 (e) and (f) are the Voronoi volumes of the same slices of Figs. 1 (c) and (d), respectively, where the $O_{CB}$ clusters are highlighted with dash line circles. No density difference exists between the $H\text{-}O_{CB}$ clusters and the surrounding melt, while the $O\text{-}O_{CB}$ domains show clearly higher density than that of melt. The above results demonstrate that nucleation of PE at quiescent condition is indeed a two-step process with $H\text{-}O_{CB}$ local structure order (LSO) as the precursor, which does not couple with density fluctuation. For the convenience to compare with nucleation under flow later, we name the two-step nucleation of PE at quiescent condition as "LSO fluctuation assisted nucleation".

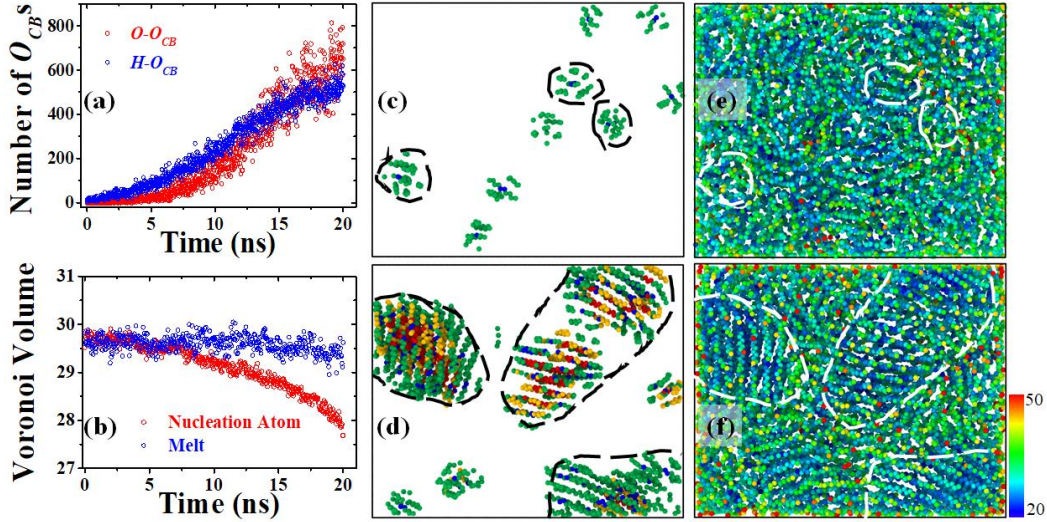

**Fig. 1.** (a) The evolutions of $O$-$O_{CB}$ (red) and $H$-$O_{CB}$ (blue) structures counted with their atom numbers. (b) The evolutions of Voronoi volume of melt (blue) and nucleation (red) atoms. (c) and (d) show $H$-$O_{CB}$ (green and blue) and $O$-$O_{CB}$ (red and yellow) clusters in a slice of systems at different time. (e) and (f) are Voronoi volumes corresponding to (c) and (d), respectively.

**Shear condition.** With the same system at quiescent, we study nucleation induced by shear flow with strain rate of 0.5 ns$^{-1}$ (Weissenberg number $W_i{\cong}25$) for 10 ns (strain of 5). To identify the effects of shear and temperature on nucleation, the systems are sheared at high temperatures $T_s$ and then quenched to 390 K for crystallization. The evolutions of $O_{CB}$ structures under these procedures are presented in Figs. 2(a)-(c) with shear temperatures $T_s$ are 400, 450 and 500 K, respectively. The red shadows cover the shear stage and the blue shadows correspond to the NPT process after quenching to 390 K. To follow density evolution during shear, Voronoi volumes of melt and nucleation atoms are plotted vs shear time in Figs. 2 (a')-(c'). At 400 K (Fig. 2 (a) and (a')), $H$-$O_{CB}$ structures (blue) form with the same Voronoi volume as that of melt at the beginning of shear, while $O$-$O_{CB}$ nuclei (red) emerge after about 3 ns of shear, at which the Voronoi volume of nucleation atoms drops deviated from melt. This process is similar with that at quiescent condition, indicating that here FIN is also a two-step nucleation assisted with LSO fluctuation, which may be due to that 400 K is lower than the melting temperature ($T_m \cong 420$ K). Nevertheless, comparing Fig. 2 (a) and Fig. 1(a) shows that imposing shear does accelerate nucleation.

At shear temperature $T_s$ = 450 K, $H$-$O_{CB}$ structures can still form but negligible $O$-$O_{CB}$ one emerges during shear (Fig. 2 (b)), while neither $H$-$O_{CB}$ nor $O$-$O_{CB}$ forms during shear at $T_s$ of 500 K. Nevertheless, after quenched to 390 K, sharp increases of both $H$-$O_{CB}$ and $O$-$O_{CB}$ contents occur, indicating flow does enhance nucleation at these two temperatures. As no $H$-$O_{CB}$ structures form at 500 K, here FIN is not LSO assisted nucleation as that at quiescent. Comparing the evolutions of Voronoi volume of melt and nucleation atoms during shear at 450 and 500 K (Fig. 2 (b') and (c'), nucleation atoms exhibit lower Voronoi volume than that of melt after shearing for about 3.6 ns, indicating density fluctuation is induced by shear. Thus under flow condition, nucleation may be assisted by density fluctuation rather than LSO.

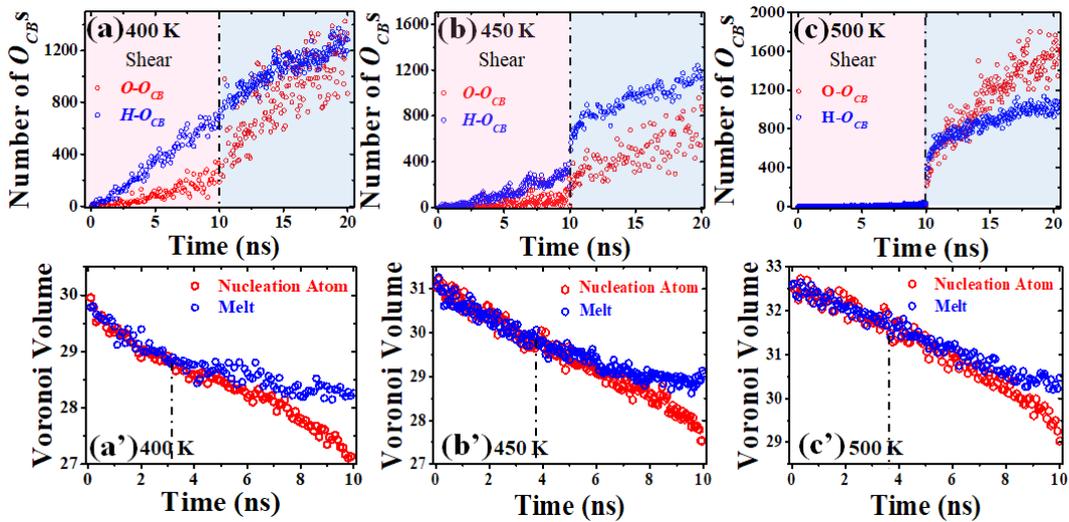

**Fig. 2.** (a) - (c) The evolutions of $O$-$O_{CB}$ (red) and $H$-$O_{CB}$ (blue) structures counted with their atom numbers. The pink shadows cover the shear stage and blue shadows correspond to NPT processes after quenched to 390 K. The shear temperatures $T_s$ are labeled in upper left corner. (a') - (c') are evolutions of Voronoi volumes of melt (MA, blue) and nucleation (NA, red) atoms during shear.

To verify whether density fluctuation assisted FIN, we closely check how nucleation takes place in the system sheared at $T_s$ of 500 K. Fig. 3 (a) is a representative slice of the sheared system just after quenched to 390 K ($t$=10.05 ns), at which nuclei of $O$-$O_{CB}$ crystal form. Note for better view, carbon atoms in $H$-$O_{CB}$ (blue and green) and $O$-$O_{CB}$ (red and yellow) clusters are shown while atoms from melt are omitted here. Fig. 3 (b) shows the Voronio volume of the same slice just after sheared at 500 K before quench ($t$=10 ns), where neither $H$-$O_{CB}$ nor $O$-$O_{CB}$ structures form yet. As shown by the Voronio volume, density distributes heterogeneously after shear. Comparing Figs. 3 (a) and (b), one can find that nucleation at 390 K exactly occurs in the domains with higher density (lower Voronoi volume) after shear at 500 K as highlighted by the dash line circles. This demonstrates that FIN is indeed assisted by density fluctuation.

To further elucidate the structure of the high density domains induced by shear, we introduces a parameter $CO$ coupling conformational and orientational orders, as flow can induce intra-chain conformational ordering and align them in parallel. $CO$ parameter is defined as the following equations:

$$P(\theta) = \frac{3\cos^2\theta - 1}{2} \tag{1},$$

$$CO = l^2 \times (2P(\theta) + 1) \tag{2},$$

where $l$ is the length of all-*trans* segments (counted with number of carbon atoms). $P(\theta)$ is the orientation parameter and $\theta$ is the angle between segments **R** and shear direction. The higher $CO$ value corresponds to a longer length and higher orientation of a conformational ordered segment (COS), while their spatial distribution represents concentration. With the same slice of the system in Fig. 3(b), we calculate *CO* and present in Fig. 3 (c), in which gray are atoms in coil state while all-*trans* atoms are colored according to their *CO* values with blue and red referring to 0 and 200, respectively. The regions of COS with high *CO* values are circled out with dash lines, which exactly correspond to the high density regions in Fig. 3(b). Evidently, here density fluctuation can be attributed to the coupling between conformational and orientational orderings induced by flow. Comparing the positions of $O$-$O_{CB}$ nuclei (Fig. 3(a)), high density (Fig. 3 (b)) and high *CO* value (Fig. 3(c)) regions, we reach a conclusion that FIN is indeed assisted by density fluctuation, which is a result of the coupling between flow-induced conformational and orientational orderings.

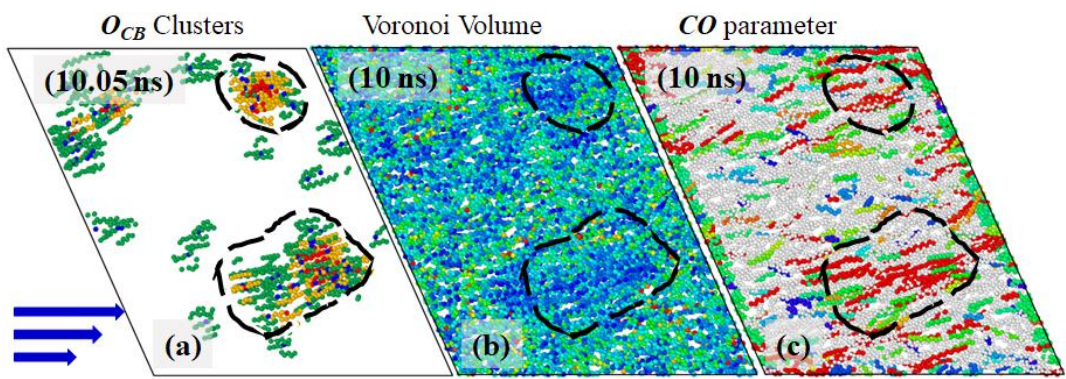

**Fig. 3.** A slice of the simulation box with thickness equals to 30 nm at $T_s$ = 500 K was taken as an example with time located at the left-upper corner. The $O_{CB}$ clusters, voronoi volume and *CO* parameter are shown in (a), (b) and (c) respectively.

To further explore the physical mechanism of the flow-induced density fluctuation, we calculate the evolutions of entropy, intra-chain and inter-chain energies (see APPENDIX) of melt and nucleation atoms during shear. Fig. 4(a) presents the entropy reduction $\Delta S$ of the whole system, which is calculated with equation:

$$T\Delta S = T\Delta S_{conformation} + T\Delta S_{orientation} = -\left(\Delta F_{elastic} + \Delta F_{orientation}/c\right)$$

$$= -\left[\frac{3k_BT}{2Ne} < \frac{l_t}{l_0} >^2 + \frac{1}{2c}\left(\boldsymbol{D} + \boldsymbol{D^T}\right):\boldsymbol{\sigma}\right] \tag{3},$$

$Ne$ is the entanglement length of PE, which is about 68 from tube model [33]. $l_0$ and $l_t$ are the end-to-end distance of the $Ne$ segments at shearing time of 0 and $t$, respectively, which are extracted from the simulation system during shear. $k_B$ is the Boltzmann constant and $T$ is temperature, $\frac{1}{2}\left(\boldsymbol{D} + \boldsymbol{D^T}\right)$ and $\boldsymbol{\sigma}$ are the strain and stress tensors respectively and $c$ is the number of segments in unit volume. $\Delta S$ shows a decrease of 0.05 $k_BT/atom$ in the early stage of shear, which is mainly due to the shear-induced orientation. Then it drops down to 0.15 $k_BT/atom$ sharply and discontinuously at $t$ of about 3.6 ns and then follows a continuously weak decrease. The discontinuous reduction of $\Delta S$ suggests that a first-order like stretch- induced coil-stretch transition occurs at 3.6 ns. After the transition, the long COS starts to grow as shown in Fig. 4(a), where the number of COS with length $\geq$ 30 is presented. Figs. 4(b) and 4(c) compare the intra-chain $E_{intra}$ and the inter-chain $E_{inter}$ energies of atoms in melt and nucleation regions, which reflect the content of *trans* conformation and the cooperative effect of chain segments, respectively. $E_{intra}$ of nucleation atoms starts to deviate from that of melt at about 3.6 ns, indicating the content of *trans* conformation in nucleation region becomes higher than that in matrix melt, which is coincidence with the occurrence of density fluctuation (see Fig. 2 (c')). Whilst obvious deviation of inter-chain $E_{inter}$ between nucleation region and melt occurs later at about 8 ns, which may correspond to the occurrence of actual phase separation.

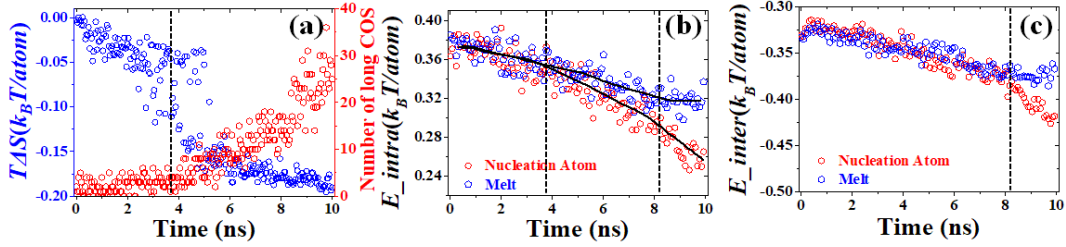

**Fig. 4.** (a) The evolution of entropy (blue) and the number of long conformational ordered segments ($l \geq 30$, red). (b) and (c) are the evolutions of Intra- and inter-chain energies of nucleation (NA, red) and melt (MA, blue) atoms, respectively.

## DISSCUTION

The above results demonstrate that two-step nucleation does occur in PE crystallization but with different intermediate orders at quiescent and flow. At quiescent, as isolated single long COS is hard to form due to entropic penalty, the cooperative effect of neighboring short *trans* segments with rotational symmetry becomes the best way to minimize the free energy and promote conformational ordering, which proceeds via *H-O$_{CB}$* LSO. The density of *H-O$_{CB}$* structures is comparable with melt and they are dynamical with small size and short life time. As soon as *H-O$_{CB}$* clusters grow to certain size, *O-O$_{CB}$* nuclei emerge inside *H-O$_{CB}$* clusters, which is promoted by coalesce of nearby *H-O$_{CB}$* clusters. Thus at quiescent condition the two-step nucleation of PE is assisted by LSO fluctuation, during which *H-O$_{CB}$* structures serve as the precursor. Under flow condition, the successive *trans* segments (long COS) can form forcedly by stretch even without cooperative inter-chain interactions, which makes nucleation following different kinetic pathways from that at quiescent. Coupling between conformational and orientational orders at flow leads to the formation of high density domains, which eventually transform into crystal nuclei. Thus the kinetic pathway of FIN actually follows a multi-stage process, namely conformational/orientational

ordering – density fluctuation – nucleation, which is fundamentally different from that at quiescent.

The above results are not only consistent with our early simulation at quiescent but also in line with experimental observations. With spectroscopic and other techniques, *trans*-rich structures or coil-helix transition are observed before the onset of crystallization of PE, isotactic polypropylene (iPP) and other polymers at quiescent [1,34,35]. Considering the entropic penalty of conformational ordering of individual segment, experimental observation on the formation of *trans*-rich structures may partially support the occurrence of LSO like $H\text{-}O_{CB}$ structures, which, however, requires further experiment to confirm. As Olmsted et al [21] suggested, density fluctuation requires high concentration of COS, which may not be fulfilled at quiescent (especially at low supercooling) but may be realized under flow. Density fluctuation prior to crystallization is well documented in the study of FIC [1,23,29,36]. Combining in-situ infrared, SAXS and WAXS techniques, recently we observed that FIC of iPP follows a multi-stage process: conformational/orientational ordering – density fluctuation – nucleation[37], which is well in line with current simulation, confirming the validity of the multi-stage nucleation model at flow.

Different kinetic pathways of nucleation at quiescent and flow challenges current models of FIN, such as the most well-recognized ERM as well as the modified ones. These approaches may phenomenologically describe the general trend of FIN but loss the essentially physical mechanism. Our simulation shows that a sharp drop of entropy occurs due to chain stretch, which promotes the growth of long COS and density fluctuation and eventually leads to phase separation to mediate nucleation. The orders of magnitude increase of nucleation rate at flow should be mainly attributed to different kinetic pathways at quiescent and flow. In this sense, we are calling to build a quantitative theory for the multi-stage FIN model with the kinetic pathway of conformational/orientational ordering – density fluctuation – nucleation, which may eventually lead to fully understanding of FIC.

In conclusion, current all-atom molecule dynamic simulation reveals that nucleation of PE takes different kinetic pathways at quiescent and flow, although both follow two-step nucleation approaches. Quiescent nucleation is mediated via LSO, while FIN goes through a multistage process via conformation/orientation - density fluctuation – nucleation, which may account the orders increase of nucleation as comparing at quiescent. The two-step nucleation models are different from Hoffman-Lauritzen model at quiescent and Flory's entropic reduction model at flow conditions, but consistent with experimental observations with spectroscopic and X-ray scattering techniques, which is also in-line with nucleation models proposed for spherical atoms and small molecules.

## ACKNOWLEDGEMENTS

The authors would like to thank Prof. Daan Frenkel (Cambridge) and Prof. Stephen Cheng (Akron) for fruitful discussion on simulation and data interpretation. National Supercomputing Center of University of Science and Technology of China and National Supercomputer Center in Tianjin (TianHe-1(A)) are acknowledged for providing the computational resources. This work is financially supported by National Natural Science Foundation of China (51633009) and also in part supported by the China Postdoctoral Science Foundation (Grant No:2016M602015).

## APPENDIX A: SIMULATION DETAILS

Full-atom MD simulations are carried out with LAMMPS packages to keep conformation and stereo-hindrance effect of PE. The OPLS_AA force field is chosen with the parameters proposed by Jorgensen [38]. The system contains 32 PE chains with 500 monomers/chain, so there are about 100,000 atoms in the simulation box. Initial structure of amorphous PE is generated by random walk using Materials Studio packages [39]. After long time relaxing at 600 K to create PE melt with $<R^2>/<Rg^2> = 5.20\pm1.45$ (mean squared end vector $<R^2>$ over radius of gyration $<Rg^2>$) then quenched down to 375 K to run dynamics for 20 ns for simulation at quiescent condition. Whilst for shear condition the system was sheared to 5 strain along the xy plane with a strain rate of 0.5 ns-1 at $T_s = 400$, 450 and 500 K and then quenched down to 390 K to run dynamics for 20 ns (the data shown in Figs. 2(a)-(c) only first 10 ns). We keep 1 atm. isobaric condition in y and z directions and isothermal during the shear while all other simulations are NPT ensemble with 1 atm. The time step is 1 fs. The periodic boundary condition is imposed in three directions.

## APPENDIX B: ENERGY AND ENTROPY CALCULATION

### Energy Calculation

The $E_{intra}$ and $E_{inter}$ were calculated using force field of OPLS_AA and based on the trajectory of carbon atoms. The $E_{intra}$ was represented by dihedral energy, which reflects the conformation transition in the system.

$$E_{intra} \approx E_{dihedral} = \tfrac{1}{2} K_1[1+\cos(\phi)] + \tfrac{1}{2} K_2[1+\cos(2\phi)]$$
$$+ \tfrac{1}{2} K_3[1+\cos(3\phi)] + \tfrac{1}{2} K_4[1+\cos(4\phi)] \tag{A4},$$

$$E_{inter} = 4\varepsilon < \left(\frac{\sigma_0}{r}\right)^{12} - \left(\frac{\sigma_0}{r}\right)^6 > \qquad r < r_c \tag{A5},$$

where $\phi$ is the dihedral angle and $r$ is the distance between two carbon atoms, other force field parameter are listed in TABLE A1:

TABLE A1. Force field parameters

| Parameter | Value | unit |
|-----------|-------|------|
| $K_1$ | 1.7400 | Kcal/mole |
| $K_2$ | -0.1570 | Kcal/mole |
| $K_3$ | 0.2790 | Kcal/mole |
| $K_4$ | 0.0000 | Kcal/mole |
| $\varepsilon$ | 0.6600 | Kcal/mole |
| $\sigma_0$ | 3.50 | Angstroms |
| $r_c$ | 10 | Angstroms |

**Entropy Calculation**

The entropy reduction comes from stretch ($\Delta F_{elastic}$) and orientation ($\Delta F_{orientation}$) as shown in Eq. (3). The change of Helmholtz free energy of a single ideal chain was defined as $\Delta F_{elastic}$ in this work and it has the form of [40]:

$$\Delta F_{elastic} = \frac{3k_B T}{2Ne} < \frac{l_t}{l_0} >^2 \tag{A6}.$$

$\Delta F_{orientation}$ was calculated using Doi-Edward tube model [33, 41],

$$\Delta F_{orientation} = -\frac{1}{2c}\left(\boldsymbol{D} + \boldsymbol{D}^T\right) : \boldsymbol{\sigma} \tag{A7},$$

where $\boldsymbol{\sigma}$ is the stress tensor and $-\frac{1}{2}\left(\boldsymbol{D} + \boldsymbol{D}^T\right)$ is the strain tensor in the following form for simple shear field,

$$-\frac{1}{2}\left(\boldsymbol{D} + \boldsymbol{D}^T\right) = \frac{1}{2}\begin{pmatrix} 0 & \dot{\gamma} & 0 \\ \dot{\gamma} & 0 & 0 \\ 0 & 0 & 0 \end{pmatrix} \tag{A8}.$$

Here $\dot{\gamma}$ is shear rate and it was 0.5 ns$^{-1}$ in this work. Combining (S3) and (S4) we have:

$$\Delta F_{orientation} = -\frac{1}{2c}\left(\dot{\gamma}\sigma_{yx} + \dot{\gamma}\sigma_{xy}\right) = -\dot{\gamma}\sigma_{xy}/c \tag{A9}.$$

$\sigma_{xy}$ can be calculated as:

$$\sigma_{xy} = \frac{3ck_BT}{N \cdot l_0^2}\sum_{n=1}^{N-1} <(\boldsymbol{R_n} - \boldsymbol{R_{n-1}})_x (\boldsymbol{R_n} - \boldsymbol{R_{n-1}})_y > \tag{A10},$$

where $\boldsymbol{R_n}$ denotes the end-to-end vector of the nth entanglement strand.

## APPENDIX C: $O_{CB}$ PARAMETER

In order to distinguish the local ordered structures in our system, a shape descriptor

defined as $O_{CB}$ was introduced based on the concept of shape matching, which is used

to transfer the multi-dimension structure into a mathematical index or similarity metric

[42]. In this work, the $O_{CB}$ parameter could be calculated as Eqs. (A11) and (A12), and

we have clearly interpret it in our former publication [X. Tang, *et al.* Phys. Rev.

Materials 1, 073401 (2017)]. $Q_l$ in Eq. (A11) is summation of spherical harmonic function $Y_{lm}$, where $l = 4$ and $m \in [0,l]$, $\theta_{ij}$ and $\varphi_{ij}$ correspond to the polar and azimuthal angles respectively. Eq. (A12) is the average operation, where $N_b(i)$ is the number of neighboring atoms $j$ of center atom $i$ within a cut_off distance of 5.4 Å.

$$Q_l = \sum_{m=0}^{l} \left| Y_{lm}\left(\theta_{ij}, \varphi_{ij}\right) \right|^2 \tag{A11}$$

$$O_{CB} = \frac{1}{N_b(i)} \sum_{j=1}^{N_b(i)} \left(\frac{2\pi}{l+1} Q_l\right)^{1/2} \tag{A12}$$